\documentclass[reprint,
aps,superscriptaddress,twocolumn,showpacs,longbibliography]{revtex4-2}
\usepackage{graphicx,color}
\usepackage{amsmath,amsfonts,enumerate,amsthm,amssymb,bbm}
\usepackage[colorlinks=true,citecolor=blue,linkcolor=magenta]{hyperref}
\usepackage{multirow}
\usepackage{bbold}

\usepackage{braket}
\usepackage{soul}

\usepackage[caption = false]{subfig}

\usepackage{scrextend}
\usepackage{xcolor}
\usepackage{dsfont}
\usepackage{nicefrac}
\usepackage[linesnumbered,ruled,vlined]{algorithm2e}
\usepackage{quantikz}
\usepackage{apptools}
\usepackage{comment}
\mathchardef\ordinarycolon\mathcode`\:
\mathcode`\:=\string"8000
\begingroup \catcode`\:=\active
  \gdef:{\mathrel{\mathop\ordinarycolon}}
\endgroup
\usepackage{amsthm}

\usepackage{cancel}
\theoremstyle{plain}
\newtheorem{thm}{Theorem}

\theoremstyle{definition}

\theoremstyle{remark}

\usepackage{amsmath,bm}
\usepackage{todonotes}

\AtAppendix{\counterwithin{thm}{section}}
\AtAppendix{\counterwithin{corol}{section}}
\AtAppendix{\counterwithin{lem}{section}}
\AtAppendix{\counterwithin{propos}{section}}
\AtAppendix{\counterwithin{defn}{section}}
\AtAppendix{\counterwithin{rmk}{section}}
\AtAppendix{\counterwithin{ex}{section}}

\def\>{\rangle}
\def\<{\langle}

\usepackage{amsmath, amssymb}

\usepackage{lipsum}
\usepackage{xcolor}
\usepackage{tikz}
\usepackage{mathtools,amsfonts,amssymb,amsthm}
\usepackage{graphicx}
\definecolor{ocre}{RGB}{52,177,201}

\begin{document}
%\title{Quantum Carleman Linearization for Stable Fluid Flows}
\title{Stability of nonlinear dissipative systems with applications in fluid dynamics}
%\title{Quantum Carleman Linearization for Stable Fluid Flows}

\date{January 2026}

\author{Javier Gonzalez-Conde}
\email[Corresponding author: ]{\qquad javier.gonzalezc@ehu.eus}

\affiliation{Department of Physical Chemistry, University of the Basque Country UPV/EHU, Apartado 644, 48080 Bilbao, Spain}
\affiliation{EHU Quantum Center, University of the Basque Country UPV/EHU, Apartado 644, 48080 Bilbao, Spain}

\author{Daniel Isla}
\affiliation{Department of Physical Chemistry, University of the Basque Country UPV/EHU, Apartado 644, 48080 Bilbao, Spain}
\affiliation{Basque Center for Applied Mathematics (BCAM), Alameda de Mazarredo 14, 48009 Bilbao, Spain}

\author{Sergiy Zhuk}
\affiliation{IBM Quantum, IBM Research Europe - Dublin, IBM Technology Campus, Dublin, Ireland}

\author{Mikel Sanz}

\affiliation{Department of Physical Chemistry, University of the Basque Country UPV/EHU, Apartado 644, 48080 Bilbao, Spain}
\affiliation{EHU Quantum Center, University of the Basque Country UPV/EHU, Apartado 644, 48080 Bilbao, Spain}
\affiliation{Basque Center for Applied Mathematics (BCAM), Alameda de Mazarredo 14, 48009 Bilbao, Spain}

\begin{abstract}
Nonlinear partial differential equations are central to physics, engineering, and finance. Except in a limited number of integrable cases, their solution generally requires numerical methods whose cost becomes prohibitive in high-dimensional regimes or at fine resolution. Nonlinear phenomena such as turbulence are notoriously difficult to predict because of their extreme sensitivity to small variations in initial conditions, except when certain stability conditions are fulfilled. Indeed, stability allows us to achieve reliable approximate dynamics, since it determines whether small perturbations remain bounded or are amplified, potentially leading to markedly different long-term behavior. Here, we investigate the stability of dissipative partial differential equations with second-order nonlinearities. By analyzing the time evolution of solution norms in Sobolev spaces, we establish a sufficient condition for stability that links the characteristics of the linear dissipative operator, the quadratic nonlinear term, and the external forcing. The resulting criterion is expressed as an explicit inequality that guarantees stability for a wide range of initial conditions. As an illustration, we apply the framework to fluid-dynamical models governed by nonlinear partial differential equations. In particular, for the Burgers equation, the condition admits a natural interpretation in terms of the Reynolds number, thereby directly linking the stability threshold to the competition between viscous dissipation and inertial advection. We further demonstrate the scope of the approach by extending the analysis to the KPP–Fisher and Kuramoto–Sivashinsky equations.
\end{abstract}

\maketitle

\textit{Introduction--} Obtaining sufficiently accurate solutions for the evolution of nonlinear dynamical systems is typically computationally expensive. However, given a nonlinear differential problem, a numerical scheme, a set of computational resources, and a simulation time, the ability to predict whether the simulation results will achieve the required level of accuracy is of fundamental importance. \cite{khalil2002nonlinear, strogatz2018nonlinear, slotine1991applied, hirsch2003differential}. This is where the concept of stability in partial differential equations (PDEs) becomes crucial. It ensures that nearby trajectories remain close under small perturbations—that is, such perturbations are not amplified—thereby reducing the required computational resources \cite{Markus1960GlobalSC, OLIVEIRA2022112676, grigorian2021globalsolvabilitystabilityoscillation}, as proven in Brudno's theorem \cite{brudno1983entropy}.

 In engineering, stability underpins the design of reliable control systems, ensuring, for example, that autopilots and robotic arms perform correctly under a broad range of operating conditions \cite{aastrom2021feedback,ogata2010modern}. In biology, it provides a framework for modeling population dynamics that settle into steady patterns over time \cite{hofbauer1998evolutionary,murray2007mathematical,may2001stability}. In economics, it supports models in which markets or broader systems return to equilibrium despite shocks or fluctuations \cite{fisher1983disequilibrium,benhabib1994indeterminacy}. Likewise, in fluid-flow dynamics, stability is essential for understanding how velocity fields evolve and whether disturbances in flows, such as turbulence, dissipate or persist over time \cite{drazin2004hydrodynamic, aerodynamic, Hydraulic, pipeline} and to be able to provide reliable state estimates and predict the future states~\cite{zhuk2023detectability,yeo2024reducing}.

\begin{figure}[b!]
    \centering
    \includegraphics[width=.5\textwidth]{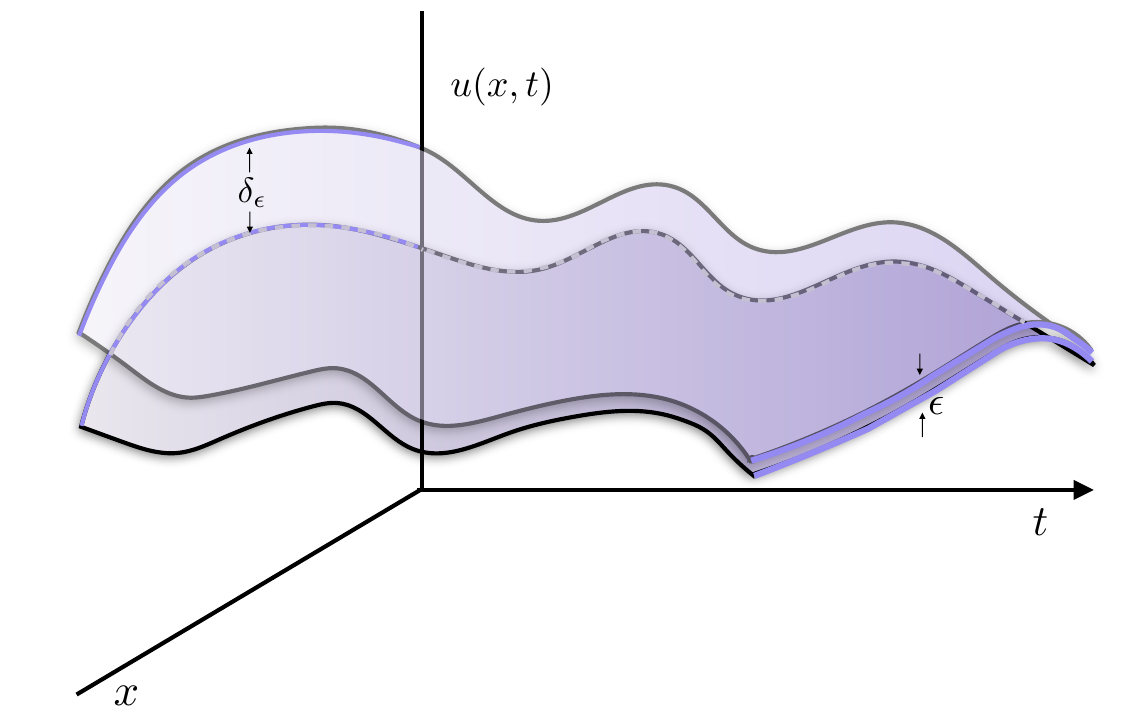}
    \caption{Stability ensures that if the initial profile is slightly altered, then the resulting evolution remains close to the reference solution for all later times, guaranteeing that numerical solutions behave consistently with the true physical problem over time. Without stability, even accurate or consistent schemes can produce meaningless results.}
    \label{fig:1}
\end{figure}

In this work, we investigate the stability properties of dissipative second-order nonlinear PDEs. In particular, we derive a sufficient stability condition expressed in terms of the parameters of the PDE. This condition is obtained by analyzing the evolution of the Sobolev norm associated with two nearby initial conditions in the Sobolev space $H^2(\Omega)$. We then examine how this stability criterion can be applied to relevant PDEs in fluid dynamics, establishing connections with key physical quantities such as the Reynolds number. \\ \\

 %\note{An alternative idea could be include a the L's in the bkl}
\textit{Theoretical framework —}Let us start by considering the family of ``generalized'' advection–diffusion nonlinear-quadratic PDEs with the form 
\begin{align}\label{Eq:Non_lienar_PDE_formulation}
    \partial_tu(x,t)=\mathcal{L}[u] +\mathcal{Q}[u]+ f(x,t) 
\end{align}
where \( d \) is the order of the differential operator, the term $\mathcal{L}[u]= ~ \sum_{l=0}^{d} \frac{1}{L^l} a_l \partial_x^{(l)}u(x,t)$ represents the linear operator, the nonlinearity is introduced via $ \mathcal{Q}[u]=~\sum_{0\leq k\leq l}^d  \frac{1}{L^{k+l}}b_{kl}\partial_x^{(k)}u(x,t)\partial_x^{(l)}u(x,t),$ and $f(x,t)$ is the forcing term that feeds energy into the system. Here for simplicity, we have rescaled our physical coordinate \(\hat x\in [0,L]\) to  the dimensionless coordinate $x=\frac{\hat x}{L}\in\Omega:=[0,1]$. Time remains dimensional and denoted by \(t\in[0,\infty)\). From now on, we presume that the solutions to Eq. (\ref{Eq:Non_lienar_PDE_formulation}) belong to the Hilbert–Sobolev space $H^{d+2}(\Omega)$, and we proceed to study their stability in the Hilbert–Sobolev space $H^{2}(\Omega)$. Additionally, we assume $u(0,t)=0$ and periodic boundary conditions for all the spatial derivatives, i.e. $ \partial_x^{(k)}u(x,t)\rvert_{x=0}=\partial_x^{(k)}u(x,t)\rvert_{x=L} \  \ \forall k \in \{0,\ldots,d\}\nonumber.$\\

\textit{Stability of trajectories —} To analyze the robustness of solutions to Eq.~(\ref{Eq:Non_lienar_PDE_formulation}) under small perturbations, we adopt a standard notion of stability formulated in the Sobolev space $H^2(\Omega)$. Informally, stability means that if the initial profile is slightly altered (in the $H^2$-norm), then the resulting evolution remains uniformly close to the reference solution for all later times, see Fig. \ref{fig:1}. More precisely, a solution \( u(x,t) \) of Eq.~(\ref{Eq:Non_lienar_PDE_formulation}) is said to be \emph{trajectory-based stable} in the \emph{Lyapunov} sense  if, for every \( \varepsilon > 0 \), there exists a \( \delta_\epsilon > 0 \) such that whenever the initial condition $u(x,0)$ is perturbed by less than $\delta_\epsilon$, i.e. $\|u(x,0) - \tilde{u}(x,0)\|_{H^2} < \delta_\epsilon$, the perturbed solution $\tilde{u}(x,t)$ satisfies that  $\|u(x,t) - \tilde{u}(x,t)\|_{H^2} < \varepsilon \quad \forall t \geq 0.$ Additionally, if on top of trajectory-based stable in the Lyapunov sense, we also have that $\lim_{t\rightarrow  \infty}\|u(x,t) - \tilde{u}(x,t)\|_{H^2}=0 $, then the solution is \emph{trajectory-based asymptotically stable}. \\

\normalsize
\textit{Theorem 1.---}  Let $u(x,t)$ be a solution to Eq. (\ref{Eq:Non_lienar_PDE_formulation}) such that $\mathcal{L}[\cdot]$ is self-adjoint and negative definite. Then, if
$y_0<\frac{\beta}{2\mu}$ and $\mathcal{R}:=\frac{f_{\text{max}}+\mu y_0^2}{\beta y_0}<1$, where  $\mu:=\mathcal{C}\sum_{0\leq k\leq l}^d  \frac{1}{L^{k+l}} |b_{kl}|>0$, $y_0:=\|u(0)\|_{H^2(\Omega)}$, $f_{\text{max}}=\max_t\|f(t)\|_{H^2(\Omega)}$ and $\beta:=-\ \text{sup}_{k\neq 0} \lambda_k$, $\lambda_k\in \sigma(\mathcal{L}[\cdot])$,  the solution $u(x,t)$ is \emph{trajectory-based asymptotically stable}.\\

\noindent Before outlining the proof of Theorem 1, we first establish sufficient conditions to find the regime where the $H^2$-norm of the solution $u(x,t)$ is upper bounded by a positive quantity, $y(t)$, that decreases monotonically in time. To this end, we study the temporal evolution of $\|u(t)\|_{H^2(\Omega)}$, and in particular of its square, i.e. $\frac{d}{dt} \lVert u \rVert_{H^2}^2 = 2 \left\langle u, \partial_t u \right\rangle_{H^2}$. After substituting the governing equation, it reads
\begin{align}
\frac{d}{dt} \lVert u \rVert_{H^2}^2 \leq 
2 \langle u, \mathcal{L}[u] \rangle_{H^2} 
+ 2 \left| \langle u, \mathcal{Q}[u] \rangle_{H^2} \right| 
+ 2 \langle u, f \rangle_{H^2}.\label{eq:energy_ineq}
\end{align}
We now examine each term in Eq.~\eqref{eq:energy_ineq} individually. Firstly, the linear operator \(\mathcal{L}\), whose contribution can be expressed as $\langle u, \mathcal{L}[u] \rangle_{H^2} = \sum_{l=0}^{d}  \frac{1}{L^l}a_l \left\langle u, \partial^{(l)} u \right\rangle_{H^2}$ 
To facilitate the spectral analysis, we introduce the eigenfunctions of \(\mathcal{L}[\cdot]\), which take the form $u_k(x) =~ C_k \exp\left(i {2\pi k} x\right),\ k \in \mathbb{Z}$,
with corresponding eigenvalues $\lambda_k = \sum_{l=0}^{d} a_l \left( \frac{2\pi i k}{L} \right)^{l}.$ 
The solution \( u(x,t) \) to Eq.~(\ref{Eq:Non_lienar_PDE_formulation}) may then be expanded in the basis of orthonormal eigenfunctions $\{u_k(x)\}_{\ k \in \mathbb{Z}}$, as  $u(x, t) =~ \sum_{k \neq 0} c_k(t) \, u_k(x)$, where we have ignored the identically zero eigenfunction. Thus, as $\mathcal{L}[\cdot]$ is self-adjoint and negative definite,  we have $\beta>0$ and therefore $ \langle u, \mathcal{L}[u] \rangle_{H^2}=\langle u,\sum_{k}  \lambda_k c_k  u_k  \rangle_{H^2} \leq~ -\beta \| u\|^2_{H^2}.$\

Next, we analyze how to bound the quadratic term. By using Cauchy–Schwarz inequality, we can bound  $|\langle u, \mathcal{Q}[u] \rangle_{H^2}|\leq \sum_{0\leq k\leq l}^d \frac{1}{L^{k+l}} |b_{kl}|\|u\|_{H^2} \|\partial_x^{(k)}u \, \partial_x^{(l)}u \|_{H^2}.$ We now use the fact that the Sobolev space \( H^{2}(\Omega) \), is a Banach algebra with respect to the standard Sobolev norm \( \|\cdot \|_{H^2} \) and pointwise multiplication. Therefore, given that $\|\partial^{(k)}\|\leq 1\  \forall k$,  we can upper bound $|\langle u, \mathcal{Q}[u] \rangle_{H^2}|\leq\mathcal{C}\sum_{0\leq k\leq l}^d  \frac{1}{L^{k+l}} |b_{kl}| \|u\|^3_{H^2},$  where $\mathcal{C}$ is constant that we can estimate as $\mathcal{C}\leq 33.02$, see Supplemental Material for further details. For the last term of Eq.~(\ref{eq:energy_ineq}), by using Cauchy–Schwarz one more time, we obtain $\langle u, f \rangle_{H^2}\leq~ \|u\|_{H^2}f_\text{max}$.  

\begin{figure*}
    \centering
\includegraphics[width=.9\textwidth]{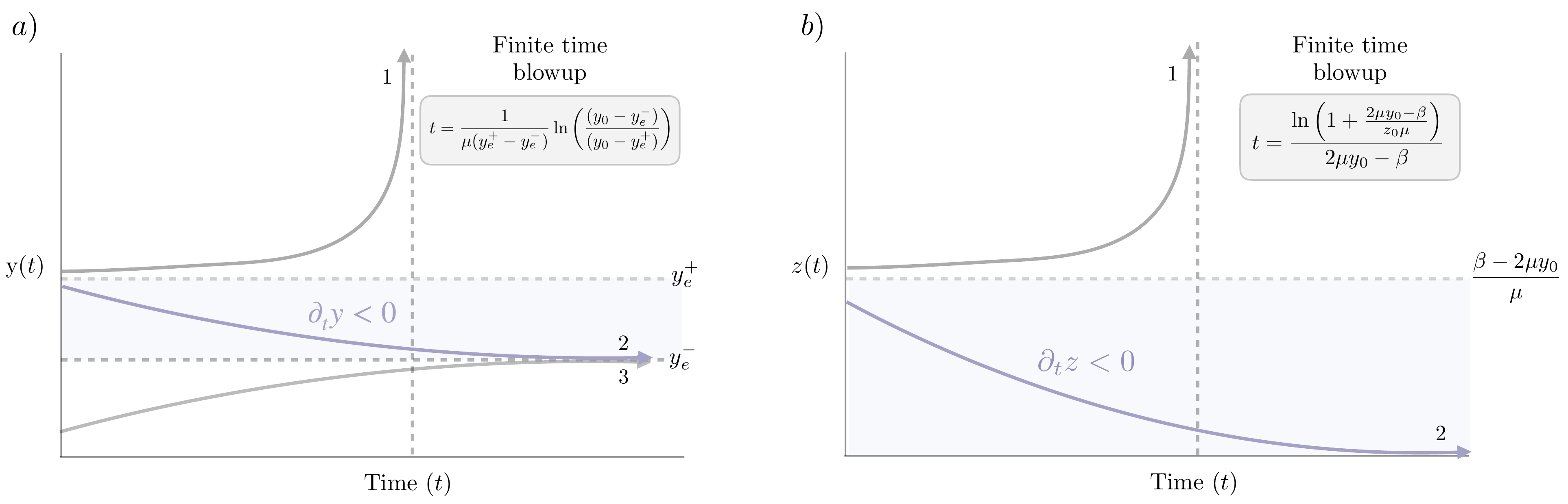}
    \caption{a) Long term behaviour of solutions to Riccati equation  with $D>0$ and different initial conditions. Solutions in region 1 blow up in finite time, while solutions in regions 2 ($\mathcal{R}<1$) and 3 are monotonically decreasing and monotonically increasing, respectively,  and both asymptotically converge to the equilibrium value $y_e^{-}$ as $t \to +\infty$. b) Long term behaviour of the Bernoulli equation when $ \mathcal{R}^*=\frac{2\mu y_0}{\beta}<1$ for  different initial conditions. Solutions in region 1 (too large initial perturbation) blow up in finite time, while solutions in region 2 (small enough initial perturbation $z_0<\frac{\beta-2\mu y_0}{\mu}$) are monotonically decreasing to zero. }
\label{fig:behaviour_riccati}
\end{figure*}

\begin{table}[t!]
\centering
\begin{tabular}{c||c}

\textbf{Condition} & \textbf{Long-term Behavior} \\
\hline
\hline
$D > 0$, $y_0 > y^+_e$ & Finite-time blowup \\
$D > 0$, $y^-_e < y_0 < y^+_e$ & Converges to $y^-_e$ $\bm{(\partial_ty<0)}$ \\
$D > 0$, $y_0 < y^-_e$ & Converges to $y^-_e$ ($\partial_ty>0$) \\
$D = 0$, $y_0 > y^*$ & Finite-time blowup \\
$D = 0$, $y_0 < y^*$ & Converges to $y^*$ ($\partial_ty>0$)\\
$D < 0$ & Finite-time blowup \\
\end{tabular}
\caption{Behaviour of solutions for different values of the discriminant $D=\beta^2-4\mu f_\text{max}$ and norm of the initial condition, $y_0$. When $D=0$, we have $y^-_e=y^+_e=y^*.$  }
\label{tab:behaviour}
\end{table}

%Therefore, gathering all the pieces, after simplifying a factor $\|u\|$ we have
%$\frac{d}{dt} \lVert u \rVert_{H^2} \leq\   \mu  \|u\|^2_{H^2}-\beta \| u\|_{H^2} +   f_\text{max}\label{eq:Riccati} $. For the sake of analyzing this inequality, we proceed to firstly study the equality, as the comparison principle establishes that the solutions of this equation will upper bound $\lVert u (t)\rVert_{H^2}\ \forall t$.  In our case, the equality leads to the Riccati equation \cite{Ricati_eq} that reads $\frac{dy}{dt} =~ \mu y^2 - \beta y + f_\text{max} $ with initial condition $y_0$. The solutions to Riccati equation are characterized by the roots of the quadratic form $Q(y) :=~ \mu y^2 -~ \beta y + f_{\text{max}}$ that we denote as $y^{\pm}_e$.

Therefore, gathering all the pieces, after simplifying a factor $\|u\|_{H^2}$, we have
$\frac{d}{dt} \lVert u \rVert_{H^2} \leq\   \mu  \|u\|^2_{H^2}-\beta \| u\|_{H^2} +   f_\text{max}\label{eq:Riccati} $.
To analyze this differential inequality, we first consider the associated equality. By the comparison principle \cite{MCNABB1988144}, the solution $y(t)$ of the equality provides an upper bound for $\|u\|_{H^2}$ for all $t$ (as long as both are defined). In our case, the equality yields the Riccati equation 
\cite{Ricati_eq}, which reads $\frac{dy}{dt} =~ \mu y^2 - \beta y + f_\text{max} $ with initial condition $y_0$. The solutions to Riccati equation are characterized by the roots of the quadratic form $Q(y) :=~ \mu y^2 -~ \beta y + f_{\text{max}}$ that we denote as $y^{\pm}_e$. These values correspond to the equilibrium solutions and depend on the discriminant $D=\beta^2-4\mu f_\text{max}$. We summarize the behaviour of the solution in every case in Table~\ref{tab:behaviour} and Fig.~\ref{fig:behaviour_riccati} ~\textcolor{magenta}{a)}. In particular, we focus on the only case in which the derivative of the solution is negative. In this case, we have $||u(t)||_{H_2}\leq y(t)<y_0 \ \forall t $. The conditions $D > 0$ and $y^-_e < y_0 < y^+_e$, can be simultaneously achieved by imposing $\mu y_0^2 -~\beta y_0 + f_{\text{max}} < 0$.  For the shake of clarity, we rewrite the condition above as 
\begin{equation}
\mathcal{R}=\frac{f_{\text{max}}+\mu y_0^2}{\beta y_0}<1.
\label{Eq:condition}
\end{equation}
\noindent We now define the characteristic effective velocity as   $v:=\frac{f_{\text{max}}+\mu y_0^2}{y_0},\ $ which represents the effective speed scale of the system. While, as shown in previous work~\cite{PhysRevResearch.7.023254}, there are cases in which, after sufficiently long times, the Kolmogorov scale induced by the forcing term becomes more refined than that induced by the initial condition, the scales defined based on this velocity avoid this issue and are guaranteed to remain constant in time.\\

In order to prove Theorem 1, we consider a perturbed initial condition $\tilde{u}(x,0)=u(x,0)+\delta(x,0)$. If we study the dynamics of the norm of the perturbation in a similar way we did with the norm of the solution and given that  $\|u(t)\|_{H_2}\leq y_0$ as $\mathcal{R}<1$, it yields 
\begin{equation}
   \partial_t\|\delta\|_{H^2} \leq \mu \| \delta\|^2_{H^2} + (2\mu y_0-\beta) \|\delta\|_{H^2}.
\end{equation}
Solving the equality (Bernoulli equation) with initial condition $z_0=||\delta(0)||_{H^2}$, we get 
\begin{equation*}
    z(t)=\frac{z_0(2\mu y_0-\beta)e^{t((2\mu y_0-\beta))}}{(2\mu y_0-\beta)+\mu z_0 (1-e^{t((2\mu y_0-\beta))})}
\end{equation*}
Consequently, if $y_0<\frac{\beta}{2\mu}$ $(\iff {\mathcal{R}^*}:=\frac{2\mu y_0}{\beta}<1)$ and we take $z_0<\frac{\beta-2\mu y_0}{\mu}$ then $\partial_t z<0$ and $z\rightarrow 0^+$ monotonically as $t\rightarrow\infty$, Fig.~ \ref{fig:behaviour_riccati}~\textcolor{magenta}{b)}, see Supplemental Material for further details. Now, given an $\epsilon>0$, if we chose the initial perturbation such that its norm is small enough, $\delta_\epsilon=||\delta(0)||_{H^2}<\min\{\epsilon,\frac{\beta-2\mu y_0}{\mu}\} $ then we have that $||\delta(t)||_{H^2}<\epsilon \ \ \forall t>0$ and furthermore as $z\rightarrow 0^+$ and $||\delta(t)||_{H^2}\leq z(t) \ \forall t$ we also obtain  $\lim_{t\rightarrow \infty}||\delta(t)||_{H^2}=0$. $\square$ \\

We now provide some relevant physical examples in fluid dynamics where we analyze the stability condition.\\

\begin{figure*}
    \centering
\includegraphics[width=.9\textwidth]{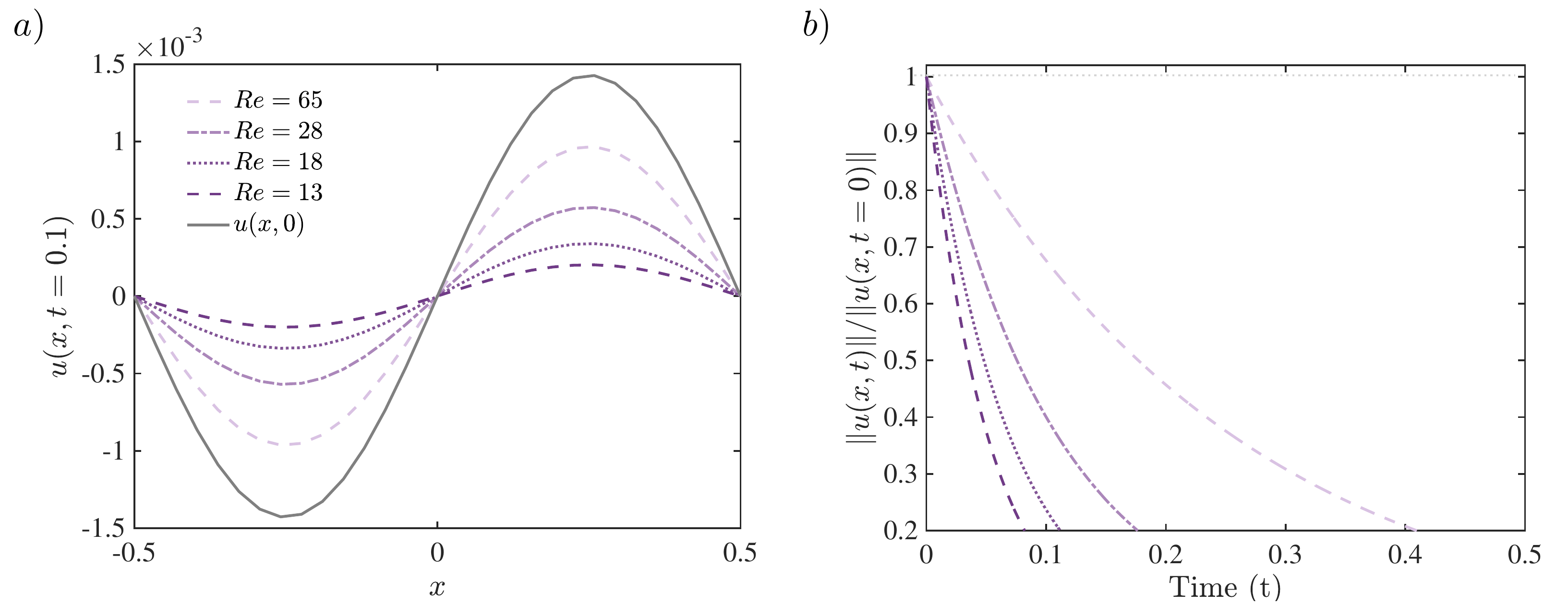}
    \caption{a) Solutions to Burgers equation for different values of the Reynolds number, $Re$, at $t=0.1$. b) Relative value in time of the norm of the solutions to Burgers equation with respect to the initial condition for different values of the Reynolds number, $Re$.}
\label{fig:Solutions_Burgers}
\end{figure*}

\textit{Burgers equation —} It captures key phenomena like convection-driven steepening, shock formation, and viscous diffusion, making it a useful toy model for studying turbulence-like behavior and wave propagation. The explicit form of the PDE reads

\begin{align}
    \frac{\partial u}{\partial t} = -  \frac{1}{L}u \frac{\partial u}{\partial x} + \frac{1}{L^2} \nu \frac{\partial^2 u}{\partial x^2} + f(x,t). 
    \label{eq:Burgers equation}
\end{align}
with $f(x,t)$ the forcing/stirring term and $\nu$ is the kinematic viscosity. Here we can identify  $\mathcal{L}_B[u]=  \frac{1}{L^2}\nu \partial_x^{2}u$. One can easily check that $\mathcal{L}_B[\cdot]$ is self-adjoint and its spectrum is given by $\bigl\{\lambda_k=-\nu \left(\frac{2\pi k}{L}\right)^2\bigl\}$, and therefore $\beta=\nu\left(\frac{2\pi }{L}\right)^2$. On the other hand 
$\mathcal{Q}_B[u]= - \frac{1}{L} u \frac{\partial u}{\partial x}$, and therefore the parameter  $\mu=\frac{\mathcal{C}}{L}$.
Gathering everything together we have that the stability conditions for Burgers equation read 

\begin{equation}
\tilde{\mathcal{R}}=\frac{vL^2}{\nu}<(2\pi)^2\label{Eq:condition_stab_1},\ \ \ \  \tilde{\mathcal{R}}^*=\frac{2\mathcal{C} y_0L}{\nu}<(2\pi)^2.
\end{equation}
In this example, the shape  of $\tilde{\mathcal{R}}$ resembles to the Reynolds number \cite{wei1989reynolds}, $Re:=U\mathcal{L}/\nu,$ where  $U$ and $\mathcal{L}$ are the characteristic velocity and length scale respectively, and $\nu$ represents the viscosity. In  particular, if $f_{\text{max}}\sim y_0^2$ and assuming $U\approx v$ then  $\tilde{\mathcal{R}}/\mathcal{L}$ can be interpreted as the Reynolds number. It indicates the ratio of the inertial to viscous forces in the fluid and its magnitude indicates whether the flow is laminar or turbulent. When the Reynolds number satisfies $Re < 1$, viscous effects outweigh inertial effects. This dominance results in greater dissipation of flow energy through viscosity, producing behavior characteristic of laminar motion. In contrast, for $Re \gg 1$, inertial forces prevail over viscous resistance, enabling the flow’s kinetic energy to foster disturbances that can develop into turbulence.
In this specific regime, we can rewrite the condition stated in Eq. (\ref{Eq:condition_stab_1}) as \begin{equation}
    Re\leq \frac{(2\pi)^2}{L} \label{Eq:condition_stab_2} %\ \ \ \ Re^*\leq \frac{(2\pi)^2}{2\mathcal{C} L} 
\end{equation}

\textit{KPP–Fisher equation —} We present a second example based on the KPP–Fisher equation, which models reaction–diffusion processes that generate traveling-wave fronts, capturing how a quantity  spreads through space over time. The PDE reads

\begin{equation}
    \frac{\partial u}{\partial t} \;=\; \frac{1}{L^2}D \, \frac{\partial^2 u}{\partial x^2} \;+\; r_1  u  - r_2u^2.
\end{equation}
 Here we can identify $\mathcal{L}_{KPP}[u]= \frac{1}{L^2}D \partial_x^{2}u + r_1u$ with eigenvalues  $\bigl\{\lambda_k=r_1-D \left(\frac{2\pi k}{L}\right)^2\bigl\}$. Therefore, in order for $\mathcal{L}_{KPP}[\cdot]$ to be negative definite, we impose a negative intrinsic growth rate, $r_1<0$, 
 and consequently $\beta=r_1-D\left(\frac{2\pi }{L}\right)^2$. This choice makes the equation describe scenarios such as biological systems in hostile environments leading to extinction, or chemical dynamics involving a decaying reactant without a source term. Regarding the nonlinear term, and $\mathcal{Q}_{KPP}[u]= -r_2u^2 $ and the parameter $\mu=\mathcal{C}r_2.$ Note that in this example $f(x,t)=0.$ \\ \\

Gathering everything together we have that the stability conditions for KPP-Fisher equation read 
\begin{equation}
    \mathcal{R}=\frac{r_2\mathcal{C}y_0}{r_1-D\left(\frac{2\pi }{L}\right)^2}<1, \ \ \ {\mathcal{R}}^*=\frac{2r_2\mathcal{C}y_0}{r_1-D\left(\frac{2\pi }{L}\right)^2}<1\label{Eq:condition_stab_2}
\end{equation}

\textit{Kuramoto–Sivashinsky —} We now study the \textit{generalized} Kuramoto–Sivashinsky equation, which models pattern formation and spatiotemporal chaos arising from the competition between destabilizing effects and stabilizing diffusion-like smoothing in dissipative fluid and flame-front systems. The equation reads
\begin{equation}
\frac{\partial u}{\partial t}=\frac{1}{L^2}a\frac{\partial^2 u}{\partial x^2}+\frac{1}{L^4}b\frac{\partial^4 u}{\partial x^4}+\frac{1}{L}cu\frac{\partial u}{\partial x}
\end{equation}
with $a,b,c\in \mathbb{R}$. Here we can identify $\mathcal{L}_{KS}[u]= \frac{1}{L^2}a\frac{\partial^2 u}{\partial x^2}+\frac{1}{L^4}b\frac{\partial^4 u}{\partial x^4}$ with eigenvalues  $\bigl\{\lambda_k=~-a\left(\frac{2\pi k }{L}\right)^2+ b\left(\frac{2\pi k }{L}\right)^4\bigl\}$. Thus, if we impose $b<0$ and $a>\frac{2\pi}{L}b$, we have $\lambda_k<~0 \ \ \forall k \neq 0$, and consequently $\beta=-a\left(\frac{2\pi  }{L}\right)^2+ b\left(\frac{2\pi  }{L}\right)^4$. With respect to to the nonlinear term we have $\mathcal{Q}_{KS}[u]= cu\frac{\partial u}{\partial x} $  and the parameter $\mu=\mathcal{C}c$. Therefore, the  stability condition for \textit{generalized} Kuramoto–Sivashinsky equation reads 
\begin{equation*}
   \mathcal{R}= \frac{c\mathcal{C}y_0}{-a\left(\frac{2\pi  }{L}\right)^2+ b\left(\frac{2\pi  }{L}\right)^4}<1,
\end{equation*} 

\begin{equation}
   \mathcal{R}^*= \frac{2c\mathcal{C}y_0}{-a\left(\frac{2\pi  }{L}\right)^2+ b\left(\frac{2\pi  }{L}\right)^4}<1 \label{Eq:condition_stab_3}.
\end{equation}

\textit{Conclusions —} In this Letter, we established a sufficient stability criterion for dissipative PDEs with second-order nonlinearities, formulated explicitly in terms of the parameters of the equation. By analyzing the evolution of the $H^2(\Omega)$ norm of both the solution and its perturbation, we showed that, under bounded quadratic nonlinearity and forcing, the solution norm decays in time and the dynamics are \emph{trajectory-based asymptotically stable}.

The resulting inequality provides a practical criterion for identifying parameter regimes in which perturbations remain bounded and solutions asymptotically converge. In the case of the Burgers equation, we further related this condition to the Reynolds number, thereby giving the criterion a clear physical interpretation in terms of the balance between inertial and viscous effects. This connection links the functional-analytic stability framework developed here to a standard dimensionless control parameter in fluid dynamics.

More broadly, the framework applies to a large class of nonlinear dissipative PDEs, suggesting potential relevance for stability analysis in more complex settings, including turbulence models and distributed-control systems in which stability constraints are essential. Our results therefore provide both a rigorous mathematical basis for stability analysis and a direct connection to physically meaningful parameters, opening the way to applications in more realistic fluid models and stability-constrained computational methods.

Natural extensions of this work include the treatment of alternative boundary conditions, non-self-adjoint dissipative operators, and the incorporation of stability-certified formulations into quantum-assisted numerical approaches for fluid-dynamical problems.

\section*{Acknowledgments} 
JGC, DI and MS thank support from the Basque Government BasQ initiative under the Q-STREAM project. They also acknowledge support from OpenSuperQ+100 (Grant No. 101113946) of the EU Flagship on Quantum Technologies, from Project Grant No. PID2024-156808NB-I00 and Spanish Ramón y Cajal Grant No. RYC-2020-030503-I funded by MICIU/AEI/10.13039/501100011033 and by “ERDF A way of making Europe” and “ERDF Invest in your Future”, from the Spanish Ministry for Digital Transformation and of Civil Service of the Spanish Government through the QUANTUM ENIA project call Quantum Spain, and by the EU through the Recovery, Transformation and Resilience Plan–Next Generation EU within the framework of the Digital Spain 2026 Agenda, and the Elkartek project KUBIBIT - kuantikaren berrikuntzarako ibilbide teknologikoak (ELKARTEK25/79).

\bibliography{Article.bib}

\clearpage
\newpage
\onecolumngrid
\section*{Supplemental Material}

\subsection{Background on Sobolev spaces $H^k(\Omega)$}

Let \( \Omega \subset \mathbb{R} \) be an open subset,  the Sobolev space \( H^{k}(\Omega) \) is defined as
\(H^{k}(\Omega) = \left\{ u \in L^2(\Omega) \; \middle| \; D^j u \in L^2(\Omega)\right.\nonumber \left.\   \forall   j  \leq 2 \right\}\), where \( D^j u \) denotes the \emph{weak derivative} of \( u \)  and \( L^2(\Omega) \) is the standard Lebesgue space of \( 2\)-integrable functions.  The Sobolev norm on \( H^{k}(\Omega) \) is given by $
\|u\|_{H^{k}(\Omega)} = \left( \sum_{j \leq k} \|D^j u\|_{L^2(\Omega)}^2 \right)^{1/2}$. Additionally \( H^{2}(\Omega) \) also defines a Hilbert space
where the scalar product reads
$  \langle u,v\rangle_{H^2(\Omega)}= \sum_{j \leq 2}\langle D^j u,D^j v\rangle_{L^2(\Omega)}$.\\ \\

Let $A:{H^{k}(\Omega)}\rightarrow {H^{l}(\Omega)}$ be a liner operator between two Sobolev spaces. We define its induced norm as $\|A\|_{k}^{l}=\sup_{u\in H^{k}(\Omega)} \frac{\|Au\|_{H^l}}{\|u\|_{H^k}}$. As an example we can consider $\partial_x:H^{1}([0,L]) \rightarrow H^{0}([0,L])$, then $\|\partial_x\|_{1}^{0} =\sup_{u\in H^{k}([0,L])}\frac{\|\partial_x u\|_{L^2}}{\sqrt{\|u\|^2_{L^2}+\|\partial_x u\|^2_{L^2}}}\leq1.$\\ \\

%\subsubsection{Multiplication in
%Sobolev spaces}
\label{Supplemental Material:Banach}
We now establish that if \( \Omega \subset \mathbb{R} \) is a bounded Lipschitz domain, then the Sobolev space \( H^{2}(\Omega) \) forms a Banach algebra with respect to the pointwise multiplication \cite{Lenka}. In other words, there exists a constant \( \mathcal{C} > 0 \) such that  $\|uv\|_{H^{2}(\Omega)} \leq \mathcal{C} \|u\|_{H^{2}(\Omega)} \|v\|_{H^{2}(\Omega)}$  \(\forall u, v \in H^{2}(\Omega) \). We now proceed to illustrate the proof of this statement.

\begin{thm} (Banach algebras in Sobolev spaces \cite{BeHo})
     The Sobolev space \( H^2(\mathbb{R}) \) is a Banach algebra with respect to the standard Sobolev norm \( \|\cdot \|_{H^k} \) and pointwise multiplication. Specifically, given \( u, v \in H^2(\mathbb{R}) \), there exists a constant $C'$ such that the following inequality holds, $\|uv \|_{H^2(\mathbb{R})} \leq C' \|u \|_{H^2(\mathbb{R})} \|v \|_{H^2(\mathbb{R})}.$
\end{thm}

\begin{thm}
(5.4 Extensions, Theorem 1 \cite{Evans}). \textit{Assume \( \Omega \) is bounded and \( \partial \Omega \) is \( C^2 \). Select a bounded open set \( V \) such that \( \Omega \subset \subset V \). Then, there exists a bounded extension linear operator $E : H^{2}(\Omega) \rightarrow H^{2}(\mathbb{R})$ such that for each \( u \in H^{2}(\Omega) \), \( Eu = u \) almost everywhere in \( \Omega \), \( Eu \) has support within \( V \) and  $\|Eu\|_{H^{2}(\mathbb{R})} \leq C \|u\|_{H^{2}(\Omega)}$ with the constant \( C \) depending only on \( V \) and \( \Omega \).}
\end{thm} 

\noindent Therefore, using Theorem 2, there exist extension operators \(E: H^{2}(\Omega) \rightarrow H^{2}(\mathbb{R})\) such that we have \( (Eu)|_{\Omega} = u, (Ev)|_{\Omega} = v  \) and \( \|Eu\|_{H^{2}(\mathbb{R})} \leq C \|u\|_{H^{2}(\Omega)}, \|Ev\|_{H^{2}(\mathbb{R})} \leq  C \|v\|_{H^{2}(\Omega)}\). Then we have the following bound \(
\|uv\|_{H^{2}(\Omega)}=  \|(Eu Ev)|_{\Omega}\|_{H^{2}(\Omega)}
\leq \|Eu Ev\|_{H^{2}(\mathbb{R})}
\leq C' \|Eu\|_{H^{2}(\mathbb{R})} \|Ev\|_{H^{2}(\mathbb{R})}
  \leq \mathcal{C} \|u\|_{H^{2}(\Omega)}  \|v\|_{H^{2}(\Omega)}.\ \ \ \square
\) \\ \\

\noindent One question of particular interest is the determination of the explicit value of the constant \( \mathcal{C}=C'C^2\). To that end, to estimate the value of the constant  $C'$, one may refer to the work presented in \cite{Morosi}, where the authors provide the numerical estimation $C'\leq 0.814$. 

We now estimate the value of $C$ through explicit extension \((Eu)(x):=\chi(x)\,\widetilde u(x), x\in\mathbb R\) for \(u\in H^2(\Omega)\) satisfying the periodic boundary conditions \(u(0)=u(1)\),\(u'(0)=u'(1)\). Here we denote by \(\widetilde u:\mathbb R\to\mathbb R\) the periodic extension of \(u\) with period \(1\),
i.e. \(\widetilde u(x)=u(x\bmod 1)\), the quintic Hermite polynomial as \(H(t)=6t^5-15t^4+10t^3, t\in\mathbb R,\) and we define the compactly supported bump \(\chi(t)=H\!\left(\frac{t+5}{5}\right)\mathbf{1}_{[-5,0]}(t)\;+\;\mathbf{1}_{[0,1]}(t)\;+\;H\!\left(\frac{6-t}{5}\right)\mathbf{1}_{[1,6]}(t)
\in C^2(\mathbb R)\) with $\mathbf{1}_{[a,b]}(t)$ the indicator function. Furthermore, denote \(M_1=\|\chi'\|_\infty=15/8\) and
\(M_2=\|\chi''\|_\infty=10\sqrt3/3\). In order to provide an explicit upper bound for the extension norm, we can estimate the norm of each
term  \((E u)=\chi\widetilde u\), \((E u)'=\chi'\widetilde u+\chi\widetilde u'\),
\((E u)''=\chi''\widetilde u+2\chi'\widetilde u'+\chi\widetilde u''\), in \(L^2(-5,6)\) by considering the inequality \(\|fg\|_{L^2}\le\|f\|_\infty\|g\|_{L^2}\) and using that
\(\widetilde u\) is \(L\)-periodic (so that \(\|\widetilde u^{(j)}\|_{L^2(-5,6)}=\sqrt11\,\|u^{(j)}\|_{L^2(\Omega)}\) for \(j=0,1,2\)),
\[\|E u\|_{H^2(\mathbb R)}^2
\le 11\Big[\left(1+2{\left(\frac{M_1}{5}\right)^2}+3{\left(\frac{M_2}{5^2}\right)^2}\right)\|u\|_{L^2(\Omega)}^2+\left(2+12\left(\frac{M_1}{5}\right)^2\right)\|u'\|_{L^2(\Omega)}^2+3\|u''\|_{L^2(\Omega)}^2\Big],
\] We can obtain the value for the constant $C$ by replacing the corresponding values of $M_1$ and $M_2$ and regrouping the terms to use the definition of the $H^2$-norm
$
\|E u\|_{H^2(\mathbb R)}^2\leq 40.5625\|u\|_{H^2(\Omega)}^2$.
Once at this stage, we can finally compute the constant $\mathcal{C}=C'C^2=0.814 \cdot 40.5625=33.0179$

\newpage

%EXTENSION BY CONVOLUTION (MOLLIFIER) EVANS' PDE BOOK  629 PAGE C.4. CONVOLUTION IN EVERY DIMENSION.

\clearpage

\clearpage
\clearpage
\clearpage

\subsection{Details on trajectory-based asymptotically stability}
%hay varios parrafos en los que hay sangrias, no se si es importante o no pero para que lo sepas, y no se si hay que meter qstream en los agradecimientos
Let  $\tilde{u}(x,t)$ and $u(x,t) $ be solutions of Eq. (\ref{Eq:Non_lienar_PDE_formulation}) such that $\tilde{u}(x,0)=u(x,0)+\delta(x,0)$, with $\delta(x,0)$ being a perturbation of the initial condition. Then, $\partial_t\delta=\mathcal{L}[\delta]+ ~\sum_{0\leq k\leq l}^d b_{kl}\partial_x^{(k)}\delta\partial_x^{(l)}u+~\sum_{0\leq k\leq l}^d b_{kl}\partial_x^{(k)}u\partial_x^{(l)}\delta.$ From here, we calculate the dynamics of the norm of the perturbation as
\begin{equation*}
    \|\delta\|_{H^2}\partial_t   \|\delta\|_{H^2}=\langle \delta,\mathcal{L}[\delta]\rangle +  \langle \delta,\mathcal{Q}[\delta]\rangle+ \langle \delta,\sum_{0\leq k\leq l}^d b_{kl}\partial_x^{(k)}\delta\partial_x^{(l)}u\rangle + \langle \delta,\sum_{0\leq k\leq l}^d b_{kl}\partial_x^{(k)}u\partial_x^{(l)}\delta\rangle 
\end{equation*}

\begin{equation*}
    \leq\langle \delta,\mathcal{L}[\delta]\rangle +  |\langle \delta,\mathcal{Q}[\delta]\rangle|+ |\langle \delta,\sum_{0\leq k\leq l}^d b_{kl}\partial_x^{(k)}\delta\partial_x^{(l)}u\rangle| + |\langle \delta,\sum_{0\leq k\leq l}^d b_{kl}\partial_x^{(k)}u\partial_x^{(l)}\delta\rangle |.
\end{equation*}
Let us now bound every term individually. First, one can easily check that $\langle \delta,\mathcal{L}[\delta]\rangle\leq -\beta \| \delta\|^2_{H^2}$ and $|\langle \delta,\mathcal{Q}[\delta]\rangle|<\mu \| \delta\|^3_{H^2}$. For the renaming two terms we have that 

   $$|\langle \delta,\sum_{0\leq k\leq l}^d b_{kl}\partial_x^{(k)}\delta\partial_x^{(l)}u\rangle|<\mu \| u\|_{H^2}\| \delta\|^2_{H^2}<\mu \| u(0)\|_{H^2}\| \delta\|^2_{H^2}.  $$ $$
    |\langle \delta,\sum_{0\leq k\leq l}^d b_{kl}\partial_x^{(k)}u\partial_x^{(l)}\delta\rangle |<\mu \| u\|_{H^2}\| \delta\|^2_{H^2}<\mu \| u(0)\|_{H^2}\| \delta\|^2_{H^2}.  $$
where we have used $\| u\|_{H^2}<\| u(0)\|_{H^2}$ as $\mathcal{R}<1$.
Therefore $\partial_t\|\delta\|_{H^2} \leq \mu \| \delta\|^2_{H^2} + (2\mu\| u(0)\|_{H^2}-\beta) \|\delta\|_{H^2}$ and
solving the equality $\partial_ty= \mu y^2_{H^2} + (2\mu y_0-\beta) y$ with initial condition $z_0$ we get 
\begin{equation}
    y=\frac{z_0(2\mu\| u(0)\|_{H^2}-\beta)e^{t((2\mu\| u(0)\|_{H^2}-\beta))}}{(2\mu\| u(0)\|_{H^2}-\beta)+\mu z_0 (1-e^{t((2\mu\| u(0)\|_{H^2}-\beta))})}.
\end{equation}
Consequently, if $(2\mu\| u(0)\|_{H^2}-\beta)<0$ then $y\rightarrow0$, and by using the comparison principle it implies that$\|\delta\|_{H^2}\rightarrow0$.

\end{document}